\documentclass[10pt, pra,twocolumn]{revtex4-1}
\usepackage{graphicx}
\usepackage{xcolor}
\usepackage{amsmath, amssymb}
\usepackage{tikz}

\let\sub\bigstar
\renewcommand{\bigstar}{\textcolor{red}{\sub}}

\usetikzlibrary{arrows,decorations.pathmorphing,backgrounds,positioning,fit,shapes.misc}

\begin{document}

\title{Systematically generated two-qubit anyon braids}

\author{Caitlin Carnahan,$^1$ Daniel Zeuch,$^2$ and N.E. Bonesteel$^2$}
\affiliation{$^1$Department of Computer Science, Florida State University, Tallahassee, Florida 32310, USA\\
$^2$Department of Physics and NHMFL, Florida State University, Tallahassee, Florida 32310, USA}

\date{\today}

\begin{abstract} 
Fibonacci anyons are non-Abelian particles for which braiding is universal for quantum computation. Reichardt has shown how to systematically generate nontrivial braids for three Fibonacci anyons which yield unitary operations with off-diagonal matrix elements that can be made arbitrarily small in a particular natural basis through a simple and efficient iterative procedure. This procedure does not require brute force search, the Solovay-Kitaev method, or any other numerical technique, but the phases of the resulting diagonal matrix elements cannot be directly controlled. We show that despite this lack of control the resulting braids can be used to systematically construct entangling gates for two qubits encoded by Fibonacci anyons.
\end{abstract}

\maketitle

\section{Introduction}
\label{intro}

In topological quantum computation, particle-like excitations that obey non-Abelian statistics are used to store and manipulate quantum information in an intrinsically fault-tolerant manner \cite{kitaev03,freedman02,nayak08}. When $N$ such particles are present, and held far enough apart, there is a degenerate Hilbert space with dimensionality exponentially large in $N$.  Distinct states in this Hilbert space cannot be distinguished by local measurements, but rather only by measurements over regions enclosing two or more particles.  When non-Abelian particles are moved around one another, so that their worldlines form braids in 2+1 dimensional space time, unitary operations are enacted on this degenerate Hilbert space.  Provided the particles continue to be kept far enough apart as they are braided, the resulting unitary operation is identical for any two topologically equivalent braids and so is robust against errors.

Fibonacci anyons are arguably the simplest non-Abelian particles for which braiding alone is sufficient to carry out arbitrary quantum computations \cite{freedman02}.  Unfortunately, these anyons appear to be much harder to realize experimentally than Majorana zero modes (see e.g., \cite{alicea12}), for which braiding is not universal for quantum computation.  Fibonacci anyons can in principle arise as quasiparticle excitations of the $k=3$ Read-Rezayi state \cite{read99}, which may describe the $\nu=12/5$ fractional quantum Hall effect, as well as the $\nu=2/3$ bosonic fractional quantum Hall effect in a rotating Bose condensate \cite{cooper01}.  More recently it has been proposed that Fibonacci anyons might be engineered in systems in which charge 2e condensates formed by clusters of fractionalized excitations in Abelian quantum Hall fluids are induced via the proximity effect with ordinary superconductors \cite{mong14,vaezi14}.  Fibonacci anyons can also appear as excitations of certain spin models \cite{levin05} and related non-Abelian surface codes \cite{koenig10_2} with the potential to one day be realized experimentally \cite{bonesteel12,wosnitzka15}. 

To use Fibonacci anyons for quantum computation, it is natural to encode logical qubits using three or four anyons with fixed topological charge \cite{freedman02}.  For these encodings, single-qubit gates can be carried out by braiding three anyons within a qubit without any leakage out of the encoded qubit space.  Braids which approximate any desired single-qubit gate can be found by carrying out brute force searches over three-anyon braids of some depth, after which either the Solovay-Kitaev method \cite{bonesteel05,hormozi09}, or the hashing technique based on finding braids that approximate the generators of the icosohedral group of \cite{burrello10,burrello11} (see also \cite{mosseri08}), can be used to systematically improve the braid, with the braid length $L$ growing as $L \sim \log^c \frac{1}{\epsilon}$ with decreasing error $\epsilon$, where $c \simeq 4$ for the Solovay-Kitaev method used in \cite{bonesteel05,hormozi09} and there is evidence that $c = 2$ for the method of \cite{burrello10,burrello11}.  More recently, using ideas from algebraic number theory, a numerical procedure for finding braids which are asymptotically optimal, i.e. for which the braid length grows as $L \sim \log \frac{1}{\epsilon}$, has been developed \cite{kliuchnikov14}.

For two-qubit gates, leakage out of the computational space will occur whenever an anyon is braided outside of its home qubit.  It is then a nontrivial problem to find braids which suppress this leakage while at the same time perform entangling two-qubit gates.  Fortunately, the problem of finding such braids for the six or eight anyons associated with two encoded qubits can be reduced to that of finding a finite number of three-anyon braids, where a strand can correspond to a single Fibonacci anyon, or a collection of anyons braided as a composite whole \cite{bonesteel05,hormozi07,hormozi09,xu08,xu09}.  This reduces two-qubit gate construction to a finite number of effective single-qubit gate constructions, which can be carried out using the methods described above.
  
Reichardt \cite{reichardt12} has shown that the braiding properties of Fibonacci anyons allow for an elegant iterative construction of three-anyon braids that carry out purely diagonal operations in certain natural bases.  Reichardt used these constructions to present a systematic procedure for distilling Fibonacci anyons from a collection of particles that could be either Fibonacci anyons or topologically trivial particles (an earlier distillation method based on brute force search was presented in \cite{koenig10_1}).  It is the purpose of the present paper to show that closely related constructions to those used in \cite{reichardt12} can be used to systematically find braids for carrying out leakage-free two-qubit gates.  The convergence of these braids is better than either those found by Solovay-Kitaev \cite{bonesteel05,hormozi07} or the icosohedral group hashing technique of \cite{burrello10,burrello11}, and is of the same order as the asymptotically optimal braids found using the procedure of \cite{kliuchnikov14}.

The paper is organized as follows.  In Sec.~\ref{computing} we review the basic properties of Fibonacci anyons and describe how to encode a logical qubit using either three or four anyons.  In Sec.~\ref{sequence} we review those aspects of Reichardt's iterative procedure which are needed to understand our two-qubit gate constructions.  Section \ref{phase} shows how braids generated by this procedure can be used in the two-qubit braid construction of \cite{hormozi09} for four-anyon qubits.  Section \ref{czb} then presents an alternate approach, closely related to that of \cite{xu08}, for constructing two-qubit braids for either three- or four-anyon qubits.  Finally, Sec.~\ref{conclusion} presents our conclusions.  

\section{Computing with Fibonacci Anyons}
\label{computing}

In the Fibonacci anyon theory \cite{freedman02,trebst08}, there are two possible topological charges: the trivial charge of the vacuum, 0, and the charge of a single Fibonacci anyon, 1.  The only nontrivial fusion rule is $1\times 1 = 0+1$, meaning that two objects (where an object can be a single anyon or a collection of anyons) with topological charge 1 can have a total topological charge of either 0 or 1.  The other fusion rules follow from the fact that $0$ is the trivial charge: $0\times 0 = 0$, $0\times 1 = 1 \times 0 = 1$.  One consequence of these fusion rules is that the Hilbert space degeneracy of $N$ Fibonacci anyons with total topological charge 0 is given by the ($N-1$)st Fibonacci number.

In addition to the fusion rules, the essential data needed to compute the unitary operations produced by braiding Fibonacci anyons is contained in two $2\times 2$ matrices, $R$ and $F$.  

The $R$ matrix for Fibonacci anyons is 
\begin{eqnarray}
R = \left(\begin{array}{cc}
e^{-i4\pi/5} & 0 \\ 0 & e^{i3\pi/5}	
\end{array}\right).
	\label{R}
\end{eqnarray}
This matrix gives the phase factors produced by moving two Fibonacci anyons around one another as shown in the following fusion tree diagram,
\begin{equation}
\def\paircoordinates{\coordinate (A) at (-.3,0);\coordinate (B) at (.3,0);\coordinate (AB) at (0,-.4);\coordinate (C) at (0,-.75);}
\begin{tikzpicture}[scale=1,baseline=-13pt]
\paircoordinates
\draw (A) -- (AB);
\draw (B) to (AB);
\draw [ultra thick] (AB) -- (C);
\node at (.2,-.6) {$b$};
\end{tikzpicture}
\rightarrow
\begin{tikzpicture}[scale=1,baseline=-13pt]
\paircoordinates
\draw [out=-150,in=150,looseness=1.8] (B) to (AB);
\draw [draw=white,double=black,double distance=.4pt,very thick,out=-30,in=30,looseness=1.8] (A) to (AB);
\draw [ultra thick] (AB) -- (C);
\node at (.2,-.6) {$b$};
\end{tikzpicture}
=
R_{bb}
\begin{tikzpicture}[scale=1,baseline=-13pt]
\paircoordinates
\draw (A) -- (AB);
\draw (B) to (AB);
\draw [ultra thick] (AB) -- (C);
\node at (.2,-.6) {$b$};
\end{tikzpicture}.
\end{equation}
Here the unlabeled thin lines all carry charge 1, while the labeled thick lines can carry either charge 0 or 1. The above diagram shows that if two Fibonacci anyons are interchanged once with a particular sense, their wavefunction acquires a phase factor of $e^{-i4\pi/5}$ if their total topological charge (labeled $b$ in the diagram) is 0, and $e^{i3\pi/5}$ if their total topological charge is 1.  If the anyons are exchanged with the opposite sense the phase factors acquired are the complex conjugates of those given above.

The $F$ matrix for Fibonacci anyons is 
\begin{eqnarray}
F = \left(\begin{array}{cc}
\phi^{-1} & \phi^{-1/2} \\ \phi^{-1/2} & -\phi^{-1}	
\end{array}\right)
	\label{F}
\end{eqnarray}
where $\phi = (\sqrt{5}+1)/2$ is the golden mean.  This matrix describes a change of basis corresponding to different ways of combining the topological charge of three Fibonacci anyons, as depicted in the following fusion tree diagram,
\begin{equation}
\def\threecoordinates{\coordinate (A) at (-.5,0);\coordinate (B) at (0,0);\coordinate (C) at (.5,0);\
\coordinate (AB) at (-.25,-.25);\coordinate (BC) at (.25,-.25);\
\coordinate (ABC) at (0,-.5);\coordinate (D) at (0,-.75);}
\begin{tikzpicture}[scale=1,baseline=-13pt]
\threecoordinates
\draw (A) -- (AB) -- (B);
\draw (C) -- (ABC);
\draw [ultra thick] (AB) -- (ABC);
\draw (ABC) -- (D);
\node at (-.3,-.5) {$b$};
\end{tikzpicture}
=
\sum_{b' \in \{0,1\}}
F_{b^\prime b}
\begin{tikzpicture}[scale=1,baseline=-13pt]
\threecoordinates
\draw (B) -- (BC) -- (C);
\draw (A) -- (ABC);
\draw [ultra thick] (ABC) -- (BC);
\draw (ABC) -- (D);
\node at (.3,-.5) {$b'$};
\end{tikzpicture}
\end{equation}
where, again, unlabeled thin lines are all assumed to carry the charge 1 and the labeled thick lines can carry either charge 0 or 1.

\begin{figure}[t]
	\includegraphics[width=\columnwidth]{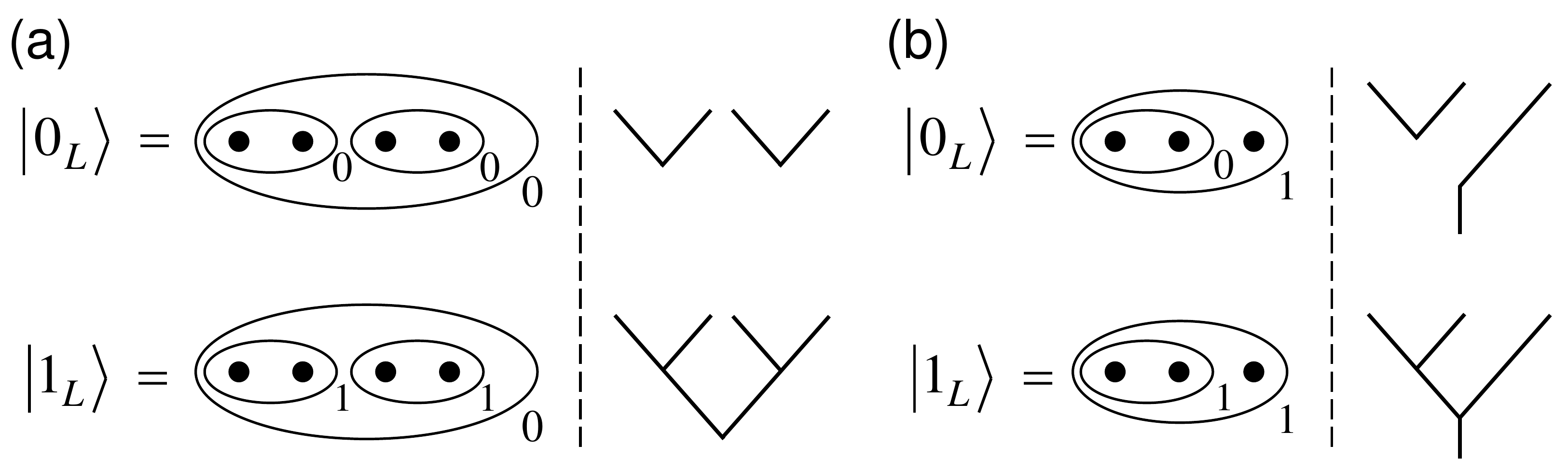}
	\caption{(a) Three-anyon and (b) four-anyon qubit encoding.  The logical qubit states $|0_L\rangle$ and $|1_L\rangle$ are shown in both oval and fusion tree notation.}
	\label{qubits}
\end{figure}

Figure \ref{qubits}(a) shows the 0 and 1 states of a logical qubit encoded using four Fibonacci anyons with total charge 0, using fusion tree diagrams as well as an alternate notation in which ovals enclose collections of particles and are labeled by the total topological charge of the enclosed particles.  In the text, we will represent states in the oval notation by replacing ovals with parentheses.  The qubit states shown in Fig.~\ref{qubits}(a) are then $|0_L\rangle  = ((\bullet \bullet)_0(\bullet\bullet)_0)_0$ and $|1_L\rangle = ((\bullet \bullet)_1(\bullet\bullet)_1)_0$ where $\bullet$ denotes a Fibonacci anyon.  A similar qubit encoding using three anyons with total charge 1 is shown in Fig.~\ref{qubits}(b).  

Single-qubit operations are carried out by braiding Fibonacci anyons within a given qubit.  The unitary operation produced by such a braid can always be determined using the $R$ and $F$ matrices.   Braiding objects within a given collection of objects will not change the total topological charge of the collection, and so there is no danger of leakage errors out of the encoded qubit space when carrying out such single-qubit gates.  Braids which carry out a desired single-qubit operation can be found by a combination of brute force search over braids up to a given length and the Solovay-Kitaev algorithm \cite{dawson06,hormozi07}, as well as other numerical methods \cite{burrello10,burrello11,kliuchnikov14}.

Finding braids for two-qubit gates is more difficult.  When anyons from two distinct qubits are braided, there will inevitably be leakage out of the encoded qubit space.  Note that any two-qubit braid that enacts an operation on a pair of three-anyon qubits [Fig.~\ref{qubits}(b)] will enact the same operation on a pair of four-anyon qubits [Fig.~\ref{qubits}(a)] .  This is because a three-anyon qubit, which has total charge 1, is equivalent to a four-anyon qubit, which has total charge 0, with one anyon removed, i.e. $((\bullet \bullet)_a(\bullet\bullet)_a)_0 = (((\bullet \bullet)_a\bullet)_1\bullet)_0$ with $a=0$ or 1. 

\section{Reichardt Sequence and Weaving}
\label{sequence}

The following observation due to Reichardt \cite{reichardt12} plays a central role in our systematic two-qubit gate constructions.  Starting with any unitary $2\times 2$ matrix, $U_0$, after each iteration of the equation
\begin{eqnarray}
	U_{k+1} = U_k R U_k^{\dagger} R^3 U_k R^3 U_k^\dagger R U_k,
	\label{iterate}
\end{eqnarray}
the magnitude $x_k$ of the off-diagonal matrix elements of $U_k$ is greatly reduced (provided $x_k\neq1$), with
\begin{eqnarray}
	x_{k+1} = x_k^5.
	\label{reduce}
\end{eqnarray}
This reduction is due to the geometric fact that the sequence of operations (\ref{iterate}) when viewed as rotations automatically cancels up to 4$^{th}$ order any deviations $U_k$ may have from being a pure $z$-axis rotation \cite{reichardt05}.  

Our two-qubit gates are built out of braids that are constructed iteratively using (\ref{iterate}).  We focus on weaves --- braids in which only one particle (or object), the weft, is mobile while the other particles, the warp, remain fixed. It has been shown that any unitary operation that can be carried out by braiding can be carried out by weaving \cite{simon06}.

To construct these weaves we follow a procedure similar to that used by Reichardt \cite{reichardt12}.  We begin by setting $U_0$ equal to a ``seed'' product of $R$ and $F$ matrices which has the form
\begin{equation}
	F R^{n_s} F R^{n_{s-1}} \cdots F R^{n_2}F R^{n_1}F.
	\label{FRF}
\end{equation}
Iterating (\ref{iterate}) will then result in a sequence of matrices, $U_k$, which become diagonal in the $k \rightarrow \infty$ limit.  Each $U_k$ in this sequence will also be expressed as a matrix product of the form (\ref{FRF}).  

\begin{figure}[t]
	\includegraphics[width=\columnwidth]{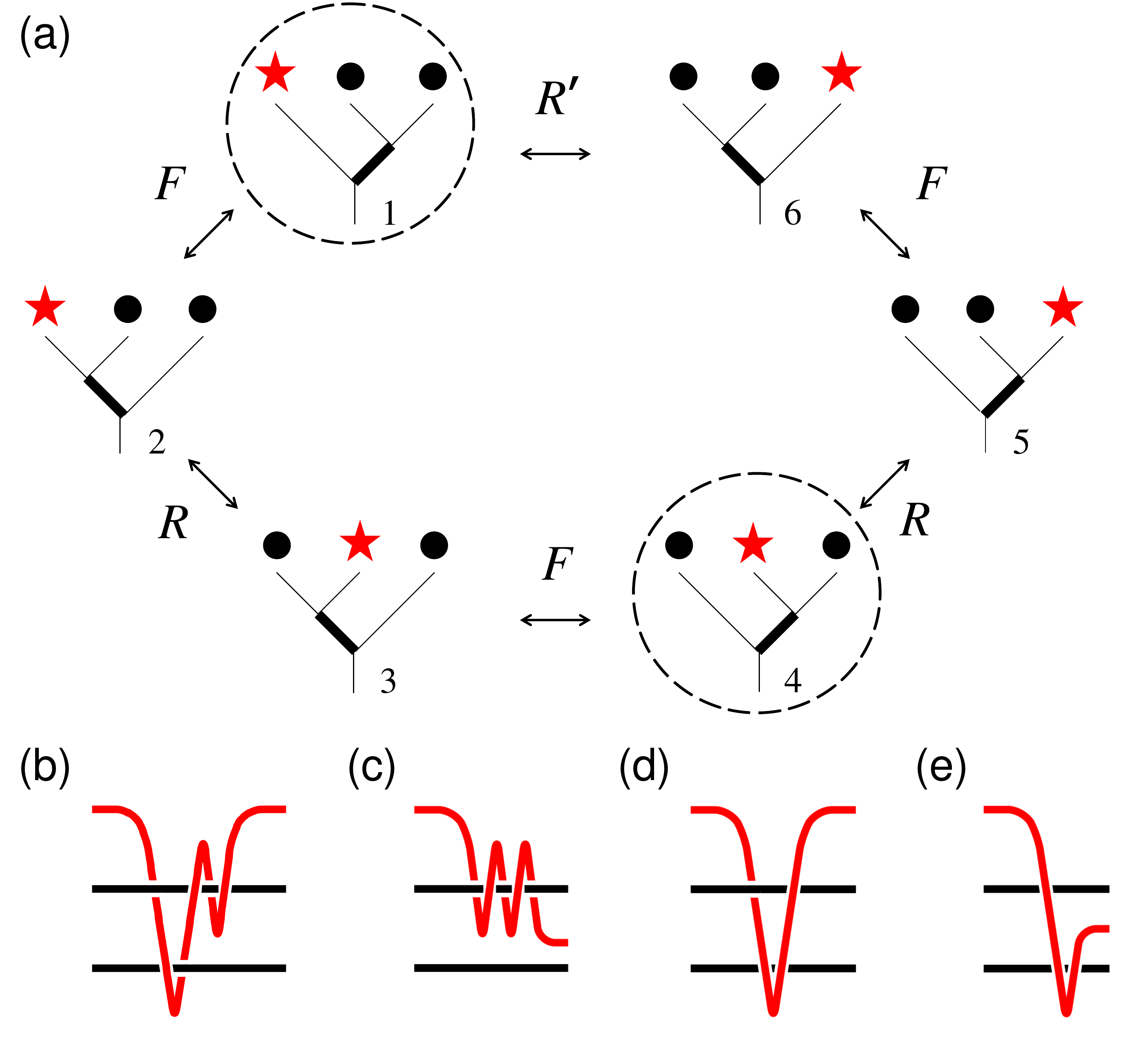}
	\caption{(color online) (a) Hexagon diagram used to find weaving patterns which correspond to matrix products of the form (\ref{FRF}).  (b) Phase weave corresponding to $F R^3 F R^{-2} F R F$.  (c) Exchange weave corresponding to $FR^{5}F$.  (d) Phase weave corresponding to $FR^{-1}FRF$ with an $R^{-1}$ operation added (resulting in $R^{-1}FR^{-1}FRF$).  (e) Exchange weave corresponding to $F$ with two $R$ operations added (resulting in $RFR$).}
	\label{hexagon}
\end{figure}

To convert products of $F$ and $R$ matrices into weaves we use the hexagon diagram shown in Fig.~\ref{hexagon}(a).  The fusion diagrams, labeled 1 through 6, at the vertices of this hexagon are given in two different bases, $\def\threecoordinates{\coordinate (A) at (-.5,0);\coordinate (B) at (0,0);\coordinate (C) at (.5,0);\
\coordinate (AB) at (-.25,-.25);\coordinate (BC) at (.25,-.25);\
\coordinate (ABC) at (0,-.5);\coordinate (D) at (0,-.75);}
\begin{tikzpicture}[scale=0.5,baseline=-7pt]
\threecoordinates
\draw (A) -- (AB) -- (B);
\draw (C) -- (ABC);
\draw [ultra thick] (ABC) -- (AB);
\draw (ABC) -- (D);
\end{tikzpicture} 
\text{ and }
\begin{tikzpicture}[scale=0.5,baseline=-7pt]
\draw (B) -- (BC) -- (C);
\draw [ultra thick] (BC) -- (ABC);
\draw (ABC) -- (A);
\draw (ABC) -- (D);
\end{tikzpicture}$, and we refer to $\def\threecoordinates{\coordinate (A) at (-.5,0);\coordinate (B) at (0,0);\coordinate (C) at (.5,0);\
\coordinate (AB) at (-.25,-.25);\coordinate (BC) at (.25,-.25);\
\coordinate (ABC) at (0,-.5);\coordinate (D) at (0,-.75);}
\begin{tikzpicture}[scale=0.5,baseline=-7pt]
\threecoordinates
\draw (B) -- (BC) -- (C);
\draw [ultra thick] (BC) -- (ABC);
\draw (ABC) -- (A);
\draw (ABC) -- (D);
\end{tikzpicture}
$
as the standard basis.  The matrix operations $U_k$ of all weaves that we use to construct two-qubit gates will become diagonal in this standard basis.  

For example, let us turn the product $F R^3 F R^{-2} F R F$ into a weave.  As will always be the case in what follows, we start at fusion diagram 1, circled in the upper left corner of Fig.~\ref{hexagon}(a), and weave the weft, denoted $\bigstar$, around the two warp particles, each denoted $\bullet$.  The first (rightmost) matrix in our example product is $F$.  Applying $F$ moves us one step around the hexagon in a counterclockwise sense to diagram 2, resulting in a basis change.  Next we apply $R$, which takes us one more counterclockwise step around the hexagon to diagram 3 by weaving the weft once around the central warp.  This is followed by another $F$ which takes us yet one more counterclockwise step around the hexagon to diagram 4, resulting in another basis change.  Next we apply $R^{-2}$, and, in this case, because there are an even number of $R$ operations, the weft weaves around the rightmost warp twice (with the opposite sense of the initial $R$ operation) and we return to diagram 4.  Thus, rather than progressing around the hexagon, after applying $R^{-2}$ (and whenever there are an even number of $R$ operations) we remain at the same fusion diagram.  The next $F$ then takes us one step around the hexagon in a {\it clockwise} sense to diagram 3, resulting in a basis change.  This is followed by $R^3$, and because the weft now weaves three times around the leftmost warp it does not return to its original position; thus we move one more clockwise step around the hexagon to diagram 2.  The final $F$ operation returns us to the original fusion diagram 1 from which we began.  The resulting weave, shown in Fig.~\ref{hexagon}(b), carries out the operation $FR^3FR^{-2}FRF$ in the standard basis.  

As a second example, consider the product $F R^5 F$.  Following the same procedure as above, the first operation $F$, then $R^5$, and finally $F$ again will move us one counterclockwise step each around the hexagon to diagram 4, circled in the lower right corner of Fig.~\ref{hexagon}(a).  As shown in Fig.~\ref{hexagon}(c), in this process we have woven $\bigstar$ five times around the central warp.  

The two weaves discussed above are examples of the two kinds of weaves we use to construct braids for two-qubit gates.  When turning a product of $F$ and $R$ matrices into a weave, we progress around the hexagon in Fig.~\ref{hexagon}(a) starting from the circled diagram 1.  For the first kind of weaves, which we call \emph{phase weaves}, the final diagram is the same as the starting diagram, and the weft returns to its original position [see Fig.~\ref{hexagon}(b)].  For weaves of the second kind, which we call \emph{exchange weaves}, the final diagram is the circled diagram 4 and the weft exchanges its position with the central warp [see Fig.~\ref{hexagon}(c)].  

For the first two examples, we could have constructed the corresponding weaves by noting that in the standard basis $R$ and $FRF$ are the elementary braid matrices for interchanging the two rightmost particles and the two leftmost particles, respectively.  However, if we followed the above procedure for the product $FR^{-1}FRF$ we would have gone nearly all the way around the hexagon of Fig.~\ref{hexagon}(a), ending at fusion diagram 6.  This diagram is not in the standard basis, and so we need to perform an additional odd number of $R$ operations to return to diagram 1.   Since our goal is to produce weaves that carry out diagonal operations in the standard basis, and $R$ itself is diagonal in this basis, we are always free in our constructions to multiply any product of the form (\ref{FRF}) on the right or left by any power of $R$.  The supplemented sequence $RFR^{-1}FRF$ then takes us back to the starting diagram 1.  Note that this last operation is denoted $R^\prime$ in Fig.~\ref{hexagon}(a).  We use this notation to distinguish this operation from $R$ because it is based on the weave represented by the following fusion tree diagram,
\begin{equation}
\def\threecoordinates{\coordinate (A) at (-.5,0);\coordinate (B) at (0,0);\coordinate (C) at (.5,0);\
\coordinate (AB) at (-.25,-.25);\coordinate (BC) at (.25,-.25);\
\coordinate (ABC) at (0,-.5);\coordinate (D) at (0,-.75);}
\begin{tikzpicture}[scale=1,baseline=-13pt]
\threecoordinates
\draw (A) -- (AB) -- (B);
\draw (C) -- (ABC);
\draw [ultra thick] (ABC) -- (AB);
\draw (ABC) -- (D);
\node at (-.3,-.5) {$b$};
\end{tikzpicture}
\rightarrow
\begin{tikzpicture}[scale=1,baseline=-13pt]
\threecoordinates
\draw [out=-30,in=25,looseness=1.2](A) to (ABC);
\draw [draw=white,double=black,double distance=1.6pt,thick,out=225,in=150,looseness=2] (BC) to (ABC);
\draw (B) -- (BC);
\draw (C) -- (BC);
\draw (ABC)-- (D);
\node at (.3,-.45) {$b$};
\end{tikzpicture}
=
R^\prime_{bb}
\begin{tikzpicture}[scale=1,baseline=-13pt]
\threecoordinates
\draw (B) -- (BC) -- (C);
\draw [ultra thick] (BC) -- (ABC);
\draw (ABC) -- (A);
\draw (ABC) -- (D);
\node at (.3,-.5) {$b$};
\end{tikzpicture}.
\end{equation}
Here
\begin{eqnarray}
	R^\prime = \left(\begin{array}{cc} 1 & 0 \\ 0 & e^{-i3 \pi/5} \end{array} \right) = e^{i4\pi/5} R,
	\label{rprime}
\end{eqnarray}
thus $R$ is equal to $R^\prime$ up to an overall phase and so can be used when iterating (\ref{iterate}).  Note that the sense used to define the $R^\prime$ exchange is the opposite of that used to define $R$, and that after carrying out an odd number of $R^\prime$ operations there will also be an associated change of basis.  The phase weave which results for this third example sequence, $RFR^{-1}FRF$, is shown in Fig.~\ref{hexagon}(d).  

Similarly, matrix products of the form (\ref{FRF}) with final diagrams 2, 3 or 5 can be turned into exchange weaves by multiplying odd numbers of $R$ matrices on the left and/or right.  Specifically, if the final diagram is 2 or 3 we multiply by an odd number of $R$ operations from the right and so start progressing around the hexagon in a clockwise rather than counterclockwise sense.  Due to the symmetry of the hexagon, the new final diagram will then be either 4 or 5.  Furthermore, any product of $F$ and $R$ matrices with final diagram 5 needs to be multiplied by an odd number of $R$ operations from the left in order to proceed to diagram 4.  In what follows we always choose the power of the additional $R$ operations to minimize the total number of elementary interchanges of the weave.  

Our last example is the operation $F$.  In the hexagon, this simply takes us one counterclockwise step from diagram 1 to 2.  Multiplication by one $R$ operation on each side results in the product $RFR$ which takes us three clockwise steps around the hexagon from diagram 1 to 4.  Figure \ref{hexagon}(e) shows the corresponding exchange weave.

We have thus established how to generate three-anyon weaves that carry out operations whose matrix representations become diagonal in the standard basis.  Starting with a seed $U_0$ of the form (\ref{FRF}) and iterating (\ref{iterate}) results in a sequence of matrices $U_k$ which converges quickly to diagonal form.  As described above, we can turn every $U_k$, given as a product (\ref{FRF}), into a weave using Fig.~\ref{hexagon}(a).  In this process one moves around the hexagon, starting at fusion diagram 1.  If the final diagram for a given $U_0$ is diagram 1 or 6 we refer to the corresponding matrix product (\ref{FRF}) as a phase seed, and every resulting operation $U_k$ can be used as a phase weave.  Any other $U_0$ will be referred to as an exchange seed for which all resulting operations $U_k$ can be used as exchange weaves.  

The leakage error of the two-qubit braids constructed below is proportional to the size of the off-diagonal elements $x_k$ of the matrix representation of the operation $U_k$ in the standard basis. Given that $x_{k+1} = x_k^5$, and the fact that the weave length grows by a factor of $5$ with each iteration of (\ref{iterate}) (not counting the constant number of additional elementary weaves due to the $R$ operations which appear explicitly) we see that the length, $L$, of the weave grows as $L \sim \log \frac1x$ \cite{reichardt12}, which is significantly better than the polylogarithmic growth that occurs when using the Solovay-Kitaev method and is comparable to the optimal case achieved using the number-theory based methods of \cite{kliuchnikov14}.

\begin{figure*}[t]
	\includegraphics[width=\textwidth]{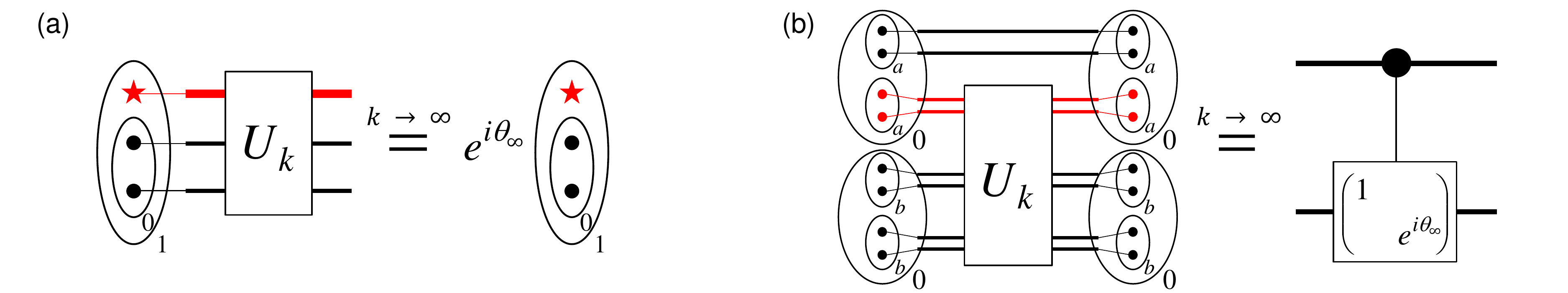}%
	\caption{(color online) (a) Box representing a phase weave, which in the $k\rightarrow\infty$ limit applies a phase factor of $e^{i\theta_\infty}$ to the state $(\bigstar(\bullet\bullet)_0)_1$. (b) Controlled-phase gate construction for two four-anyon qubits.}
	\label{phase_gate}
\end{figure*}

\section{Controlled-Phase Gates}
\label{phase}

In this section we show how the phase weaves of the previous section can be used to directly carry out entangling two-qubit gates for four-anyon qubits using a construction presented in \cite{hormozi09}.

In Fig.~\ref{phase_gate}(a), the box labeled $U_k$ with three incoming and outgoing anyon strands represents a generic phase weave which returns the weft, $\bigstar$, to its original position.  As described in the previous section, this weave is obtained by iterating (\ref{iterate}) starting from a phase seed $U_0$.  The matrix $U_k$ corresponding to the operation carried out by this weave quickly converges to a diagonal matrix, so to a good approximation its action for large $k$ is to multiply the state $(\bigstar(\bullet\bullet)_{0})_1$ by a phase factor of $e^{i\theta_k}$.  As shown in the figure, in the $k\rightarrow\infty$ limit this approximation becomes exact and the phase $\theta_k$ converges to a limiting value $\theta_\infty$.  The convergence behavior of $\theta_k$ is discussed below.

Following Hormozi et al. \cite{hormozi09}, any phase weave can be used to construct a two-qubit gate. Figure \ref{phase_gate}(b) shows two four-anyon qubits in states $a$ and $b$ and a braiding pattern for the bottommost six anyons.  To construct this pattern, neighboring pairs of anyons are first grouped into three objects, one with total charge $a$ in the upper qubit and two with total charge $b$ in the lower qubit, as indicated in the figure.  The three-anyon weave shown in Fig.~\ref{phase_gate}(a) is then carried out as a ``superweave" with the charge $a$ object as the weft and the two charge $b$ objects as the warp.

The two-qubit gate character of the resulting operation becomes evident when considering different two-qubit states defined by $a$ and $b$.  Note that any anyon weave that returns the weft to its original position results in the identity operation if either the weft or each warp object has charge zero.  The operation $U_k$ thus trivially enacts the identity if $a=0$ or $b=0$.  In the nontrivial case $ab=11$ the superweave carries out the phase weave of Fig.~\ref{phase_gate}(a) and for $k\rightarrow\infty$ applies the phase factor $e^{i\theta_\infty}$.  As shown in Fig.~\ref{phase_gate}(b), the two-qubit braid thus enacts a controlled-phase gate which for nontrivial phases $\theta_\infty \ne 0$ (mod $2\pi$)  is entangling.

\begin{figure*}[t]
	\includegraphics[width=\textwidth]{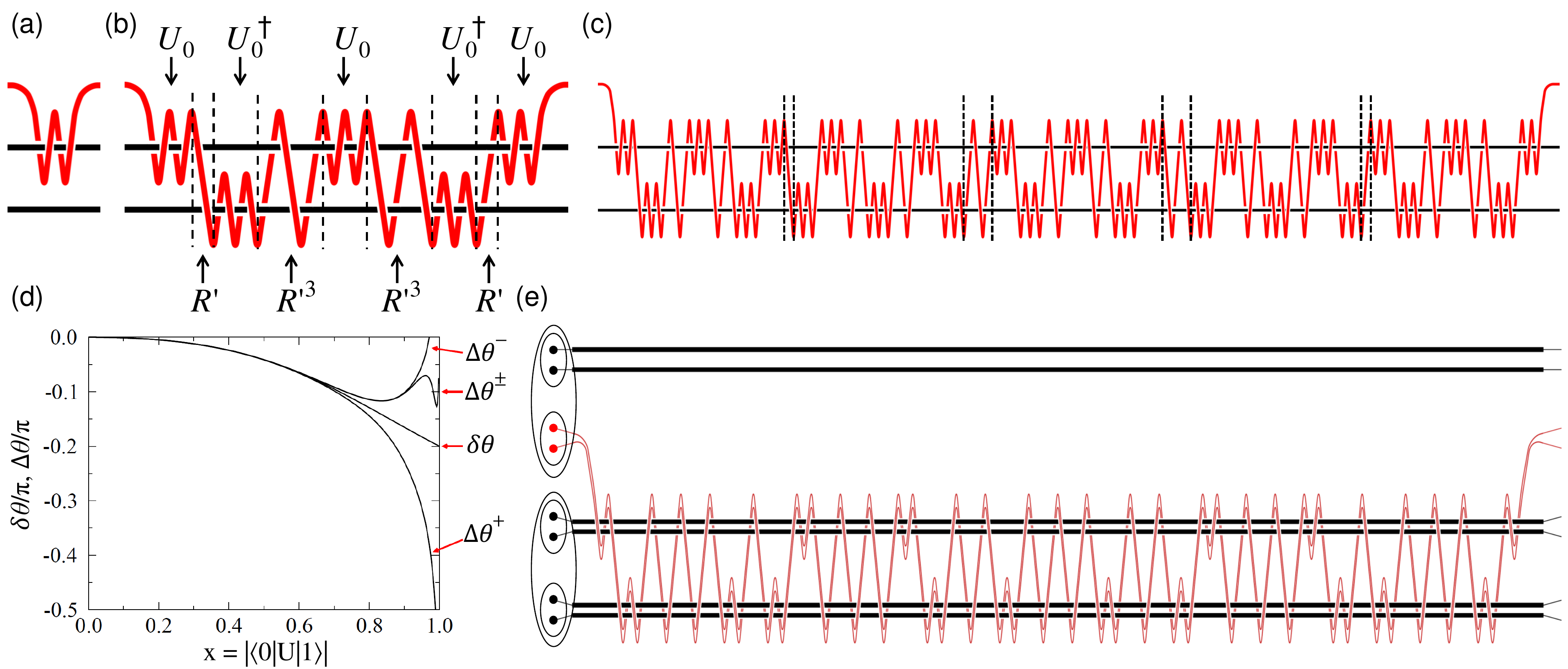}
\caption{(color online) (a) Weave corresponding to phase seed $FR^4F$.  (b) Weave obtained with this seed after one iteration of (\ref{iterate_prime}).  (c) Weave obtained with this seed after a second iteration of (\ref{iterate_prime}).  In (b) and (c) dashed lines divide parts of the weave corresponding to $U_k$ and $U_k^\dagger$ from those corresponding to the $R^\prime$ and ${R^\prime}^3$ operations appearing explicitly in (\ref{iterate_prime}).  For the seed $FR^4F$ the magnitude of the initial off-diagonal matrix elements is $x_0 \simeq 0.571$ and the intial phase is $\theta_0 \simeq 0.546 \pi$. After two iterations $x_2  \simeq 8.30 \times 10^{-7}$ and $\theta_2 \simeq 0.488 \pi$.  (d) $\delta \theta$, shift in the phase of $\langle 0 |U| 0 \rangle$ after one iteration of (\ref{iterate_prime}), plotted as a function of $x = |\langle 1 | U | 0 \rangle|$, as well as $\Delta \theta^+$, $\Delta\theta^-$, and $\Delta\theta^\pm$ corresponding to different choices for the signs $s_k$ in (\ref{deltatheta}).  (e) Controlled-phase gate (see Fig.~\ref{phase_gate}(b)) with $\theta_2 \simeq \theta_\infty \simeq 0.488 \pi$ obtained using the second iteration phase weave shown in (c).}
\label{phase_weave}
\end{figure*}

Figure \ref{phase_weave}(a) shows an example phase weave corresponding to the seed $F R^4 F$.  The weaves that belong to the first and second iterations of (\ref{iterate}) are shown in Figs.~\ref{phase_weave}(b) and (c).  Recall that when turning such sequences into weaves, some $R$ operations are realized as $R^\prime$, which is equal to $R$ up to an overall phase.  For the current construction we need to keep track of this overall phase because the value of $\theta_\infty$ determines the two-qubit gate of Fig.~\ref{phase_gate}.  In Figs.~\ref{phase_weave}(b) and (c) we see that all $R$ operations which appear explicitly in (\ref{iterate}) are $R^\prime$ operations.  This is always the case for phase weaves so that in the present section the iteration prescription can unambiguously be written as
\begin{equation}
	U_{k+1} = U_k {R^\prime}^{\pm 1} U_k^{\dagger} {R^\prime}^{\pm 3} U_k {R^\prime}^{\pm 3} U_k^\dagger {R^\prime}^{\pm 1} U_k,
	\label{iterate_prime}
\end{equation}
provided the starting operation $U_0$ is that obtained by turning the phase seed into a weave following Sec.~\ref{sequence}.
Note that in (\ref{iterate_prime}) we have used the fact that the sign of the powers of $R$ in (\ref{iterate}) can be changed without altering the result (\ref{reduce}). 

Denoting the two states in the standard basis 
$\def\threecoordinates{\coordinate (A) at (-.5,0);\coordinate (B) at (0,0);\coordinate (C) at (.5,0);\
\coordinate (AB) at (-.25,-.25);\coordinate (BC) at (.25,-.25);\
\coordinate (ABC) at (0,-.5);\coordinate (D) at (0,-.75);}
\begin{tikzpicture}[scale=0.5,baseline=-7pt]
\threecoordinates
\draw (B) -- (BC) -- (C);
\draw [ultra thick] (BC) -- (ABC);
\draw (ABC) -- (A);
\draw (ABC) -- (D);
\end{tikzpicture}
$
with total charge 0 or 1 of the two rightmost anyons by $|0\rangle$ or $|1\rangle$, respectively, and letting
\begin{equation}
	x_k = |\langle 1 | U_k | 0 \rangle|, \qquad \theta_k = \arg \langle 0 | U_k | 0 \rangle,
\end{equation} 
direct calculation yields
\begin{eqnarray}
	x_{k+1} = x_k^5, \qquad
	\theta_{k+1} = \theta_k + s_k \delta \theta(x_k),
	\label{iterate2}
\end{eqnarray}
where
\begin{eqnarray}
	\delta \theta(x) = - \arcsin \left(\frac{5^{1/4} (\phi^{-3/2} x^2 + \phi^{1/2} x^4)}{2 \sqrt{1+x^2 + x^4 + x^6 + x^8}}\right).
\end{eqnarray}
Here the sign $s_k=\pm1$ in (\ref{iterate2}) is equal to the sign of the powers of $R^\prime$ in (\ref{iterate_prime}).  

For the $k\rightarrow\infty$ phase we then have
\begin{eqnarray}
	\theta_\infty &=& \theta_0 + \Delta \theta(x_0),
	\label{formula}
\end{eqnarray}
where
\begin{eqnarray}
	\Delta \theta(x_0) = \sum_{k=0}^\infty s_k \delta\theta(x_0^{5^k}).
	\label{deltatheta}
\end{eqnarray}

\begin{table}[b]
\begin{tabular}{lcccc}
\hline\hline
SEED & $x_0$ & $\theta_0/\pi$ & $(\theta_0 + \Delta\theta^+)/\pi$ & $(\theta_0 - \Delta\theta^+)/\pi$\\
\hline
$FR^2F$ & 0.924 & 1 & 0.737 & -0.737 \\
$FRFR^3F$ & 0.924 & 1 & 0.737 & -0.737  \\
$FR^3FR^{-3}F$ & 0.882 & 0 & -0.207 & 0.207 \\
$FR^4F$ & 0.571 & 0.546 & 0.488 & 0.604 \\
$FR^5FR^5F$ & 0.415 & 1 & 0.997 & -0.997\\
$FRFRF$ & 0 & 0 & 0 & 0 \\
\end{tabular}
\caption{Example phase seeds.}
\label{table2}
\end{table}

The function $\delta\theta(x)/\pi$ is shown in Fig.~\ref{phase_weave}(d).  As $x\rightarrow 0$, $\delta\theta(x)$ vanishes quadratically, and as $x \rightarrow 1$, $\delta\theta(x) \rightarrow \pi/5$.  Three different results for $\Delta\theta(x)$, corresponding to different choices for the sequence of signs, $\{s_k\}$, are also shown in Fig.~\ref{phase_weave}(d).  $\Delta\theta^+$ corresponds to choosing $s_k = +1$ for all $k$, $\Delta\theta^{-}$ corresponds to choosing $s_0 = +1$ and $s_k=-1$ for all $k > 0$, and $\Delta\theta^\pm$ corresponds to choosing $s_k = (-1)^k$ for all $k$.  The Taylor expansions of all these functions agree up to 8th order in $x$ and only begin to noticeably deviate from one another for $x$ larger than $\sim 0.6$.  As a consequence, for seeds with small $x_0$ (which are desirable since they will converge faster to diagonal matrices), only the first contribution to the phase, $\delta\theta(x_0)$, will be appreciable, with $\theta_\infty \simeq \theta_0 \pm \delta\theta(x_0)$.

Figure \ref{phase_weave}(e) shows an explicit example two-qubit braid based on the weave of Fig.~\ref{phase_weave}(c).  Note that in Fig.~\ref{phase_weave}(e), we have eliminated trivial sequences of braids whenever one interchange is directly followed by its inverse.  In addition, using the fact that $R^{10} = 1$, we reduce the number of consecutive windings experience by neighboring strands to be less than or equal to five.  

Table \ref{table2} lists example phase seeds together with their $x_0$ and $\theta_0$ values.  Given a nontrivial phase weave (i.e. one for which $\theta_\infty \ne 0$ (mod $2\pi$)) one can always, as above, construct a leakage-free entangling two-qubit gate.  The first two entries with identical values show that different seeds can be effectively equivalent.  The last entry of Table~\ref{table2}, $FRFRF$, results in a weave that carries out a trivial operation for which $\theta_k = 0$ for all $k$ and is thus not useful in our two-qubit gate construction.  

In this section we described two-qubit gates whose specific operation cannot be a priori chosen because it depends on the phase weave used to construct the two-qubit braid.  In the next section we present a new construction based on exchange weaves for which the resulting operation of the entangling two-qubit gate is independent of the weave used.  

\begin{figure*}
	\includegraphics[width=\textwidth]{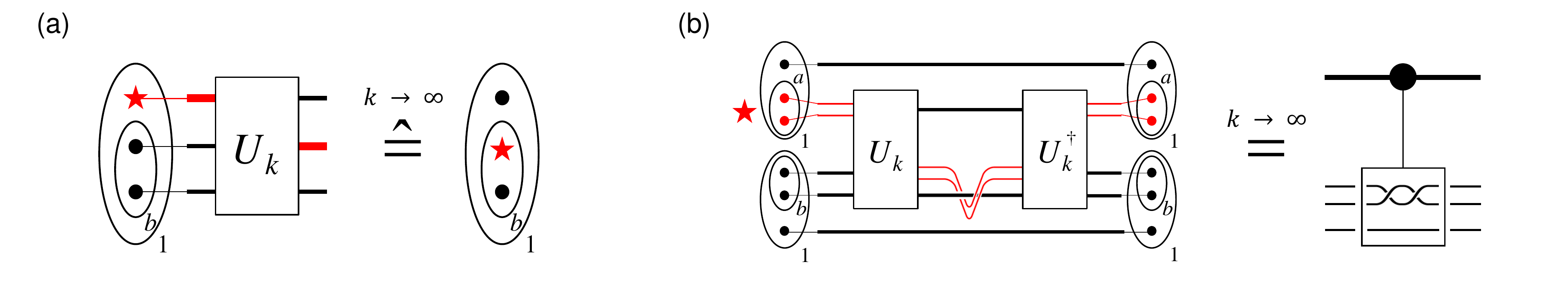}%
	\caption{(color online) (a) Box representing an exchange weave, which in the $k\rightarrow\infty$ maps $(\bigstar(\bullet\bullet)_0)_1$ to $(\bullet(\bigstar\bullet)_0)_1$. (b) Controlled-$R^2$ gate construction for two three-anyon qubits.}
\label{czbgate}
\end{figure*}

\section{Controlled-Braid Gates}
\label{czb}

We now show how the exchange weaves of Sec.~\ref{sequence} can be used to carry out two-qubit gates.  Our approach is similar to that of Xu and Wan \cite{xu08} where anyons are exchanged between qubits while essentially preserving the stored quantum information.  These exchanges, whose corresponding braids can be found via methods based on brute force search \cite{xu08}, can be performed more efficiently by the exchange weaves introduced in this paper.  One property of the two-qubit braids of \cite{xu08} is that they can only be applied to qubits encoded using four anyons each.  Here, we construct new two-qubit braids that can be applied to a pair of either three- or four-anyon qubits. 

The box labeled $U_k$ shown in Fig.~\ref{czbgate}(a) represents a three-anyon exchange weave that carries out an operation $U_k$ and is obtained, as described in Sec.~\ref{sequence}, by choosing an exchange seed $U_0$ and iterating (\ref{iterate}) $k$ times.  This exchange weave switches the positions of the weft $\bigstar$ and the central warp, while the matrix representation of $U_k$ in the standard basis converges to diagonal form for large $k$.  It follows that acting with $U_k$ on $(\bigstar(\bullet\bullet)_b)_1$, with $b=0$ or 1 in the limit of $k\rightarrow\infty$ results in the state $(\bullet(\bigstar\bullet)_b)_1$ up to a phase factor [as indicated by $\hat =$ below in the text as well as in Fig.~\ref{czbgate}(a)] whose value will be shown shortly to be irrelevant.  

Figure \ref{czbgate}(b) shows two three-anyon qubits (given in a convenient basis for what follows) and a braiding pattern for the middle four anyons $((\textcolor{red}{\bullet\bullet})_a(\bullet\bullet)_b)_d$, where $d=0$ or 1.  To construct this pattern, we first group the two anyons with total charge $a$ into a single object.  The three-anyon weave shown in Fig.~\ref{czbgate}(a) is then carried out as a superweave with this charge $a$ object as the weft and the two uppermost anyons in the bottom qubit as the warp.

To determine the action of the two-qubit weave of Fig.~\ref{czbgate}(b) first note that by the end of the weave the weft is returned to its original position.  It follows that if $a=0$ this braid carries out the identity operation.  In the non-trivial case of $a=1$ we replace the object $(\textcolor{red}{\bullet\bullet})_{a=1}$ by a single anyon, $\bigstar$,
$$
	((\textcolor{red}{\bullet\bullet})_{a=1}(\bullet\bullet)_b)_d \ \rightarrow \ (\bigstar(\bullet\bullet)_b)_d.
$$
In the limit of $k\rightarrow\infty$ the first operation shown in Fig.~\ref{czbgate}(b), $U_k$, then carries out the mapping
\begin{equation}
	\lim_{k\rightarrow \infty} U_k (\bigstar(\bullet\bullet)_b)_d \ \hat= \ (\bullet(\bigstar\bullet)_b)_d,
	\label{uk}
\end{equation}
up to a phase factor that depends on both quantum numbers $b$ and $d$.  Notice that in the case of $bd=10$ this map is already exact for any finite $k$, because the Hilbert space of three Fibonacci anyons with total charge 0 is one-dimensional.  As shown in the figure, in the next step we weave $\bigstar$ twice around the warp inside the oval with total charge $b$, thus carrying out an $R^2$ operation on the Hilbert space spanned by the states (\ref{uk}) with $b=0$ and 1.  Finally, the operation $U_k^\dagger$ exchanges $\bigstar$ with what is now the topmost warp.  Importantly, $U_k^\dagger$ further multiplies each state with given quantum numbers $bd$ by the complex conjugate of the phase factor that was applied by $U_k$.  Since the $R^2$ operation in the center of the sequence is diagonal in $b$ and $d$ the phase factors of $U$ and $U^\dagger$ cancel one another.  The resulting operation carried out by the two-qubit braid of Fig.~\ref{czbgate}(b) is then an $R^2$ operation acted on the qubit in state $b$ if $a=1$ and the identity if $a=0$.  This operation thus enacts a controlled-$R^2$ gate where the qubits in states $a$ and $b$ are, respectively, the control and the target qubit.  

The two-qubit braids of \cite{xu08}, which are applied to four-anyon qubits, are composed of a braid sequence similar to that of our construction, i.e. exchange braid---braid within the target qubit---inverse exchange braid.  These exchange braids involve all four anyons of one of the qubits and so cannot be applied to three-anyon qubits.  As opposed to the exchange braid in (\ref{uk}), in which within the $a=1$ sector different states $((\textcolor{red}{\bullet\bullet})_{a=1}(\bullet\bullet)_b)_d$ are multiplied by different phase factors, the exchange braids of \cite{xu08} multiply the entire $a=1$ two-qubit sector by a single overall phase factor.  Because of this, the braids of \cite{xu08} allow for arbitrary controlled operations while here this controlled operation needs to conserve the quantum numbers $b$ and $d$.  Accordingly, the $R^2$ operation in Fig.~\ref{czbgate}(b) can be replaced by any even power of $R$.  However, since $R^{10}=1$ this yields only four distinct entangling two-qubit gates.  

\begin{figure*}
	\includegraphics[width=\textwidth]{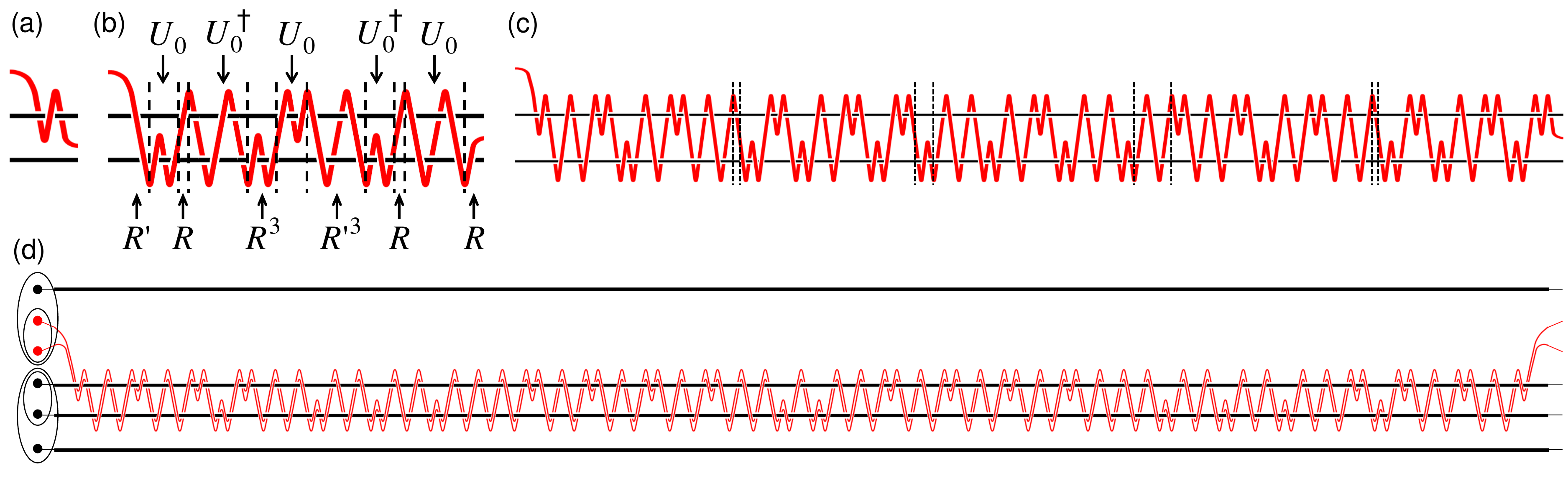}%
	\caption{(color online) (a) Weave corresponding to the exchange seed $FR^3F$.  (b) Weave obtained with this seed after one iteration of (\ref{iterate}).  (c) Weave obtained with this seed after a second iteration of (\ref{iterate}).  For the seed $FR^3F$ the magnitude of the initial off-diagonal matrix elements is $x_0 \simeq 0.300$, and, after two iterations, $x_2 = 8.67\times 10^{-14}$.  In (b) and (c) dashed lines divide parts of the weave which corresponding to $U_k$ and $U_k^\dagger$ from those corresponding to the $R$ and $R^3$ operations appearing explicitly in (\ref{iterate}).  (d) Controlled-$R^2$ gate (see Fig.~\ref{czbgate}(b)) obtained using the second iteration exchange weave shown in (c).}
\label{injection}
\end{figure*}

\begin{table}[b]
\begin{tabular}{lcc}
\hline\hline
SEED & $x_0$ & Initial Sense \\
\hline
$FR^5F$ & 0.972 & $\circlearrowleft$ \\
$F$ & $\phi^{-1/2}$ = 0.786 & $\circlearrowright$\\
$FRF$ & $\phi^{-1/2}$ = 0.786 & $\circlearrowleft$ \\
$FRFR^2F$ & $\phi^{-1/2}$ = 0.786  & $\circlearrowright$\\
$FR^3F$ &  0.300 & $\circlearrowleft$ \\
$FR^3FR^5FR^3F$ & 0.0438 & $\circlearrowright$\\
\\
\end{tabular}
\caption{Example exchange seeds.}
\label{table_czbgate}
\end{table}

When determining the operation carried out by the braid of Fig.~\ref{czbgate}(b) we made use of the crucial property that the  quantum number $a$ is conserved.  In the middle step of our construction, for $a=1$ the superweave applied to $((\textcolor{red}{\bullet\bullet})_{a=1}\bullet)_b$ is carried out with $(\textcolor{red}{\bullet\bullet})_{a=1}$ as the weft object and thus trivially conserves $a$.  However, if we instead allow for arbitrary controlled braids of these three anyons with the constraint that $a$ is conserved, then we uncover an infinite number of two-qubit entangling braids for three-anyon qubits.  One natural choice for these controlled braids are phase weaves $((\bullet\bullet)_{a=1}\bigstar)_b \rightarrow e^{i\theta_{b}}((\bullet\bullet)_{a=1}\bigstar)_b$ (to a good approximation) with the rightmost anyon as the weft $\bigstar$ and the two leftmost anyons as the warp.  [Note that the $a=0$ sector gets multiplied by an overall phase factor because here we have $((\bullet\bullet)_{a=0}\bigstar)_1 \rightarrow e^{i\theta_{a=0}}((\bullet\bullet)_{a=0}\bigstar)_1$.]  While the length $L$ of the resulting two-qubit braid then grows as $L \sim \log \frac1x$, control of the exact two-qubit operation, which depends on the phase factors applied by the controlled weave, is limited (as for the two-qubit braids of the previous section).  

We now construct an example exchange weave starting from the seed $FR^3F$.  Figures \ref{injection}(a) through (c) show the weaving patterns of the seed as well as the first and second iterations.  The sense with which one starts progressing around the hexagon in Fig.~\ref{hexagon}(a) for a given matrix product of the form (\ref{FRF}) for an exchange weave gets reversed after iterating (\ref{iterate}).  As discussed in Sec.~\ref{sequence}, whenever this sense is clockwise one needs to multiply the product on the right by an odd number of $R$ operations.  For the example, $FR^3F$, the initial sense is counterclockwise and so for the first iteration this sense is clockwise while for the second it is again counterclockwise.  We therefore multiply the first iteration by an additional $R$ operation from the right.  Since, further, the final diagram of this supplemented product is diagram 5 in the hexagon, we also multiply the sequence by one additional $R$ operation on the left.  Both additional $R$ operations are indicated in Fig.~\ref{injection}(b).  

Figure \ref{injection}(d) shows the explicit two-qubit gate braid of Fig.~\ref{czbgate} that is obtained when replacing the exchange weaves $U_k$ and $U_k^\dagger$ by the weave of Fig.~\ref{injection}(c) and its inverse, respectively.  As in the two-qubit phase gate shown in Fig.~\ref{phase_gate}(e), all redundant braids have been eliminated and the fact that $R^{10} = 1$ has been used to reduce the number of windings whenever possible.  

Table \ref{table_czbgate} lists example exchange seeds together with the magnitude of their off-diagonal elements, $x_0=|\langle0|U_0|1\rangle|$, and the sense with which one starts to move around the hexagon of Fig.~\ref{hexagon} when turning the seed into a weave.  The key parameter of any seed is the absolute value of its off-diagonal matrix elements, $x_0$, which determines how many times (\ref{iterate}) needs to be iterated to achieve a desired error.  

\section{Conclusion}
\label{conclusion}

In this paper we have shown how to use a construction introduced by Reichardt \cite{reichardt12} for distillation of Fibonacci anyons to systematically generate weaving patterns which can be used to carry out leakage-free two-qubit gates.  We presented two constructions, one, based on a two-qubit gate construction given in \cite{hormozi09}, requires four-anyon qubits.  The second is closely related to that given in \cite{xu08}, although unlike in that work, our two-qubit gates can be applied to three-anyon qubits as well as four-anyon qubits.  

Both constructions presented here are based on iterating (\ref{iterate}) to obtain weaves which produce operations diagonal in a particular basis.  The resulting weaves converge efficiently, with the length of the weave growing logarithmically in the inverse error as opposed to polylogarithmically, as is the case for the Solovay-Kitaev method.

\acknowledgments 

NEB and DZ are grateful to the JARA-Institute for Quantum Information for its warm hospitality during a visit over which part of this work was carried out.  This work was partially supported by US DOE Grant No. DE-FG02-97ER45639.

\bibliography{bibliography}

\end{document}